\documentclass[10pt, conference, letterpaper]{IEEEtran}

\usepackage[utf8]{inputenc}
\usepackage[]{cite}
\usepackage{url}
\usepackage{tabularx}
\usepackage{algorithm}
\usepackage[noend]{algpseudocode}
\usepackage{amsmath}
\usepackage{amssymb}
\usepackage{algorithmicx}
\usepackage{xspace}
\usepackage{footmisc}
\usepackage{enumitem}
\usepackage[dvipsnames]{xcolor}
\usepackage[english]{babel}

\usepackage[]{todonotes}

\def\ego{\textsc{EGO}\xspace}
\def\superego{\textsc{Super-EGO}\xspace}
\def\gdsjoin{\textsc{GDS-Join}\xspace}
\def\dbscan{\textsc{DBSCAN}\xspace}
\def\hilbert{\textsc{FGF-Hilbert}\xspace}
\def\tensor{\textsc{TED-Join}\xspace}
\def\wmmaRef{\textsc{WMMA-Ref}\xspace}

\title{Leveraging GPU Tensor Cores for Double Precision Euclidean Distance Calculations}

\author{\IEEEauthorblockN{Benoit Gallet}
\IEEEauthorblockA{benoit.gallet@nau.edu}
\IEEEauthorblockN{Northern Arizona University\\
School of Informatics, Computing, and Cyber Systems\\
Flagstaff, Arizona, USA}
\and
\IEEEauthorblockN{Michael Gowanlock}
\IEEEauthorblockA{michael.gowanlock@nau.edu}
\IEEEauthorblockN{Northern Arizona University\\
School of Informatics, Computing, and Cyber Systems\\
Flagstaff, Arizona, USA}}

\begin{document}

\maketitle


\begin{abstract}
Tensor cores (TCs) are a type of Application-Specific Integrated Circuit (ASIC) and are a recent addition to Graphics Processing Unit (GPU) architectures. As such, TCs are purposefully designed to greatly improve the performance of Matrix Multiply-Accumulate (MMA) operations. While TCs are heavily studied for machine learning and closely related fields, where their high efficiency is undeniable, MMA operations are not unique to these fields. More generally, any computation that can be expressed as MMA operations can leverage TCs, and potentially benefit from their higher computational throughput compared to other general-purpose cores, such as CUDA cores on Nvidia GPUs. In this paper, we propose the first double precision (FP64) Euclidean distance calculation algorithm, which is expressed as MMA operations to leverage TCs on Nvidia GPUs, rather than the more commonly used CUDA cores. To show that the Euclidean distance can be accelerated in a real-world application, we evaluate our proposed TC algorithm on the distance similarity self-join problem, as the most computationally intensive part of the algorithm consists of computing distances in a multi-dimensional space. We find that the performance gain from using the tensor core algorithm over the CUDA core algorithm depends weakly on the dataset size and distribution, but is strongly dependent on data dimensionality.  Overall, TCs are a compelling alternative to CUDA cores, particularly when the data dimensionality is low ($\leq{4}$), as we achieve an average speedup of $1.28\times$ and up to $2.23\times$ against a state-of-the-art GPU distance similarity self-join algorithm. Furthermore, because this paper is among the first to explore the use of TCs for FP64 general-purpose computation, future research is promising.
\end{abstract}


\section{Introduction}
\label{sec:introduction}
Tensor cores (TCs) are a type of Application-Specific Integrated Circuit (ASIC), and are specifically designed for Matrix Multiply-Accumulate (MMA) operations. The high specificity of TCs makes them typically more efficient at computing MMA operations, than other more general-purpose cores such as CPU cores or GPU CUDA cores. Given four matrices $A, B, C$, and $D$, TCs are designed to compute $D = A \times B + C$ (where $C$ and $D$ may be the same matrix). Over the past few years, TCs have been heavily used for machine learning and other fields requiring linear algebra, and few papers have examined broadening the use of TCs for other algorithms. Despite their high specificity, TCs may also be very versatile: any computation expressed with MMA operations, as defined above, should be able to leverage TCs and consequently, benefit from their high computational throughput.

Several companies have proposed a version of TCs, each with its own different characteristics. In this paper, we focus on the Nvidia GPU TCs. These TCs were first introduced with the Volta generation in 2017\footnote{\url{https://images.nvidia.com/content/volta-architecture/pdf/volta-architecture-whitepaper.pdf}}. Since this first iteration, they have been implemented in several GPU models and have greatly improved over time~\cite{nvidia_turing, nvidia_a100, nvidia_h100}. In particular, while the first generation of TCs was only capable of computing in half precision using 16-bit floats (FP16), TCs are now capable of double precision computing using 64-bit floats (FP64) with the Ampere generation~\cite{nvidia_a100}. This enables TCs to be used for applications where high precision is critical. Furthermore, their number, as well as their theoretical computational throughput, have continued to increase, making them an attractive alternative to the general-purpose CUDA cores.

As mentioned above, in this paper we focus on TCs proposed by Nvidia on their GPUs. In addition to the CUDA API to access GPU functionalities, we also leverage the Warp Matrix Multiply-Accumulate (WMMA) API~\cite{cuda_programming_guide, programming_tensor_cores}, which provides programmatic access to TCs. While other libraries also give access to TCs, they are all higher level than the WMMA API and less versatile, thus less suited to our use case. However, there are some limitations when using the WMMA API. In particular, matrix sizes are limited to a few options, and not all compute precisions are available or can be combined (e.g., FP32 for both multiplication and accumulation is not available, and FP16 multiplication can not be combined with FP64 accumulation).

The Euclidean distance is a metric commonly used in many scientific applications, particularly for data analysis algorithms such as the distance similarity self-join~\cite{damon19, superego, fgfHilbert, hegjoin}, the $k$NN~\cite{gowanlock_knn}, or the Density-Based Spatial Clustering of Applications with Noise (\dbscan)~\cite{dbscan} algorithms. Within these algorithms, distance calculations are usually the most time-consuming fraction of the total computation~\cite{dbscan_revisited}. In this paper, we propose to improve the throughput of Euclidean distance calculations by leveraging TCs on the GPU, and consequently also improve the overall performance of the algorithms mentioned above. To illustrate greater applicability to these other algorithms, we use the distance similarity self-join algorithm as a representative example case for the other data analysis algorithms mentioned above. Given a dataset $V$ in $d$ dimensions, the distance similarity self-join algorithm finds all pairs of points $(a, b)$ that are within a distance threshold $\epsilon$ of each other; $dist(a,b) \leq \epsilon$, where $a, b \in V$, and \emph{dist} is the Euclidean distance function.

This paper makes the following contributions:

    \noindent\textbullet We propose a new algorithm for computing Euclidean distances using TCs, leveraging the Nvidia Ampere architecture TCs~\cite{nvidia_a100} supporting double precision (FP64) computations.
    
    \noindent\textbullet We integrate the aforementioned method into the distance similarity self-join algorithm, that we name Tensor Euclidean Distance Join (\tensor). We show that \tensor is competitive with the best parallel distance similarity self-joins in the literature for multi-core CPUs and GPU CUDA cores. 
    
    \noindent\textbullet The solution we propose here extends beyond the distance similarity self-join algorithm and can be integrated into other algorithms that use the distance similarity self-join, or more generally Euclidean distance calculations, as a building block.
    
    \noindent\textbullet We evaluate \tensor across a broad range of datasets, that span several distributions, sizes and dimensionalities, and compare it to a state-of-the-art GPU CUDA cores (\gdsjoin~\cite{damon19}) and two multi-core CPU distance similarity join algorithms, \superego~\cite{superego} and \hilbert~\cite{fgfHilbert}. We conclude that \tensor should always be preferred over \superego and \hilbert, and should be preferred over \gdsjoin when the dimensionality $d \leq 4$, where it achieves an average speedup of $1.28\times$ (and $1.07\times$ when considering all the experiments we conducted), and up to $2.23\times$.
    
    \noindent\textbullet To our knowledge, this paper proposes the first Euclidean distance calculation for TCs using FP64 computation, and the first use of TCs for the distance similarity self-join.

The paper is outlined as follows: we present essential material in Section~\ref{sec:background}, including an overview of TCs. We then present in Section~\ref{sec:distance_calc_tensor} our solution that uses TCs to compute Euclidean distances and its integration into the distance similarity self-join algorithm. We show in Section~\ref{sec:exp_eval} the performance of our solution compared to the state-of-the-art distance similarity self-join algorithms, and we conclude and propose future research directions in Section~\ref{sec:conclusion}.


\section{Background}
\label{sec:background}
    \subsection{Problem Statement}
    \label{sec:problem_statement}
    For two points $a$ and $b$ in $d$ dimensions, and where $a_i$ represents the $i^{th}$ coordinate of the point $a$, and where $i = 1, \ldots, d$, the Euclidean distance between $a$ and $b$ is defined as follows:
    \begin{equation}
        dist(a, b) = \sqrt{\sum^{d}_{i=1}(a_i - b_i)^2}.
        \label{eq:euclideanDistance}
    \end{equation}
    The distance similarity self-join algorithm, as described above, takes a dataset $V$ in $d$ dimensions as well as a search distance $\epsilon$ as inputs, and finds all the pairs of points $(a,b)$ such that $dist(a,b) \leq \epsilon$ where $a,b \in V$, and where the distance function is, in this case, the Euclidean distance defined in Equation~\ref{eq:euclideanDistance}. For a query point $a$, finding all the other points in $V$ within $\epsilon$ from $a$ is called a range query ($|V|$ range queries in total).


    \subsection{Tensor Cores (TCs)}
    \label{sec:tensor_cores}
    TCs on GPUs are an Application-Specific Integrated Circuit (ASIC) designed for Matrix Multiply-Accumulate (MMA) operations. Given four matrices $A, B, C$ and $D$, this MMA operation is expressed as $D = A \times B + C$. Matrices $C$ and $D$ are the accumulators and may be equivalent. In hardware, TCs are designed to process $4 \times 4$ MMA operations. However, the WMMA API only gives access to larger matrices (e.g., $16 \times 16$). Therefore, several TCs are used concurrently to perform MMA operations larger than $4 \times 4$. Due to their highly specific design, TCs are significantly more efficient at MMA operations than CUDA cores: double precision computation is presented as twice as efficient when using TCs compared to CUDA cores on the Nvidia A100 GPU~\cite{nvidia_a100}. This significantly higher processing throughput is our motivation to transform Euclidean distance calculation into MMA operations, and yield higher computational throughput.

    The WMMA API~\cite{cuda_programming_guide, programming_tensor_cores} provides some low-level access to TCs, giving us the highest versatility possible. However, several limitations come along with this WMMA API. In particular, it is limited to certain matrix sizes and compute precisions. Among the options available, only a few are relevant to our work. In this paper, we focus on FP64 computation, which limits us to only one size for each of our matrices. Let $M_{m,n}$ be a matrix with $m$ rows and $n$ columns. The matrices that we can use with double precision are thus $A_{8,4}$, $B_{4,8}$, $C_{8,8}$, and $D_{8,8}$. We refer the reader to the documentation~\cite{cuda_programming_guide} for the other TCs options.

    Programmatically, the matrices proposed by the WMMA API are called \textit{fragments}, and are stored into the GPU threads registers. The WMMA API defines several functions:

        \noindent\textbullet \textit{load\_matrix\_sync()}: Load a matrix fragment from memory.
        
        \noindent\textbullet \textit{store\_matrix\_sync()}: Store a matrix fragment into memory.
        
        \noindent\textbullet \textit{mma\_sync()}: Perform an MMA operation using TCs.
        
        \noindent\textbullet \textit{fill\_fragment()}: Fill a matrix fragment with a specified value.

    As their name suggests, these function calls are synchronized. Hence, all 32 threads of the warp are blocked until the operation is complete. The \textit{load} and \textit{store} functions take, among other arguments, a stride between the elements comprising the matrix rows. Hence, all the elements consisting of a row in the target matrix need to be coalesced in memory. Furthermore, the individual elements of the matrix fragments are stored in an unspecified order in the registers. Thus, contrary to regular arrays, the first element of a matrix may not be stored in the first element of the fragment. Consequently, operations on an individual element of a matrix fragment need to be applied to all the other elements, using a loop iterating over all of the elements of the fragment.


    \subsection{Tensor Cores in the Literature}
    \label{sec:tensor_literature}
    As mentioned above, the literature concerning TCs heavily revolves around machine learning and other closely related fields, and not many other types of applications employing TCs~\cite{tensorScanReduction, tensorDbscan, tensorSearches, tensorFft, tensorDfft}. We present in this section a selection of papers that discuss the use of TCs for applications that focus more on computational/data-enabled science, similarly to this paper. Moreover, since most of the literature seems to focus on low precision computations, we believe that this paper is the first to propose an implementation using TCs for FP64 computations.
    
    Dakkak et al.~\cite{tensorScanReduction} propose a method to perform reduction and scan operations, using the WMMA API to leverage TCs. Their reduction algorithm consists of multiplying a matrix whose first row are ones and the rest are zeros with a matrix containing the values to reduce, and accumulated with a matrix containing the result from previous reductions. Their scan solution is similar but uses an upper triangular matrix filled with ones and where the rest are zeros, instead of a single row filled with ones. Their proposed solutions achieve a speedup of $100\times$ for the reduction and $3\times$ for the scan, compared to other state-of-the-art methods not using TCs.
    
    Ji and Wang~\cite{tensorDbscan} propose using TCs to improve the performance of the \dbscan algorithm. They mainly use TCs to compute distance matrices between the points that might form a cluster, using the cosine similarity formula (in contrast to the Euclidean distance used in this paper). They also use TCs to perform reductions, which are used to determine if points belong to a cluster or not. Their solution using TCs achieves a speedup of up to $2.61\times$ to compute distance matrices compared to using the CUDA cores. While this work is very relevant to us, it differs in that they use a different distance metric (cosine similarity vs. Euclidean distance), they do not use an index structure, and part of their work is exclusive to the \dbscan algorithm. In comparison, our solution essentially concerns the Euclidean distance calculations and, therefore, more applications than the distance similarity self-join that we just take as an example for this paper.
    
    Ahle and Silvestri~\cite{tensorSearches} theorize using TCs to compute similarity searches. They use TCs to compute either the Hamming, squared $L_2$ distances, or cosine similarity through an inner product operation, expressed as matrix multiplications. Additionally, they opt for the Local Sensitivity Hashing (LSH) method, reducing the overall complexity of the computation similarly to an indexing structure used by other similarity join solutions~\cite{superego, fgfHilbert, damon19}. However, and contrary to these solutions, the LSH method typically yields an approximate result.
   

    \subsection{Distance Similarity Joins}
    \label{sec:distance_simjoin}
    We discuss in this section several state-of-the-art parallel distance similarity self-join algorithms~\cite{damon19, superego, hegjoin, fgfHilbert}, which we use as reference implementations for our experimental evaluation. These selected algorithms have in common that they use an indexing structure to prune the number of distance calculations, which is a commonly used optimization~\cite{index_supported_sj, ego}. When using an index, it is first \textit{searched} to yield a set of candidate points for each query point. The set of candidate points is then \textit{refined} using distance calculations to keep pairs of query and candidate points that are within $\epsilon$ of each other.

    Kalashnikov~\cite{superego} proposes \superego, a parallel CPU algorithm to compute a distance similarity join, which is an improvement over the Epsilon Grid Order (\ego) algorithm proposed by B\"{o}hm et al.~\cite{ego}. \superego performance relies on a grid index and which is dependent on the search distance $\epsilon$, where a grid with cells of size $\epsilon \times \epsilon$ is laid on the search space to efficiently prune the candidate points to refine. Furthermore, the author proposes to reorder the dimensions of the points based on their variance, so dimensions with the highest variance are considered first when computing the distance between two points. Hence, their cumulative distance is more likely to reach $\epsilon$ sooner, allowing the short-circuiting of the distance computation, and thus to not consider the remaining dimensions. \superego has been since improved by Gallet and Gowanlock~\cite{hegjoin}, as part of a CPU-GPU distance similarity self-join algorithm. Among the changes, their version of \superego is capable of FP64 computation while performing better than \superego proposed by Kalashnikov~\cite{superego}. As such, further references to \superego in this paper will refer to the work conducted by Gallet and Gowanlock~\cite{hegjoin}, rather than Kalashnikov~\cite{superego}.

    Perdacher et al.~\cite{fgfHilbert} propose \hilbert, a parallel CPU distance similarity join algorithm also based on an epsilon grid order, but using space-filling curves as their indexing method. Using an \ego-sorted dataset, space-filling curves are used to determine, for each query point, a range of consecutive candidate points in the dataset. The authors further improve the performance by using the OpenMP API and low-level vectorized instructions, making their solution highly optimized. Because \hilbert typically performs better than \superego, particularly in higher dimensions, it is considered a state-of-the-art CPU distance similarity join algorithm. Because of some of its optimizations, \hilbert is only capable of FP64 computation.

    Gowanlock and Karsin~\cite{damon19} propose \gdsjoin, a GPU algorithm for high-dimensional distance similarity self-joins. Their optimizations related to the high-dimensional case include reordering the dimensions of the points based on their variance, so these with the highest variance would be considered first when computing distances. Similarly to \superego presented above, this particularly pairs well with distance calculation short-circuiting. Overall, dimensions with a higher variance are susceptible to increase the cumulative distance more than dimensions with lower variance and are thus more likely to trigger short-circuiting the distance calculation. They also propose to index the data in fewer dimensions than the input dataset dimensionality, making their grid index efficient even in higher dimensions, as the cost of searching their grid index is bound by the number of dimensions that are indexed. Furthermore, as their source code is publicly available, it appears that new optimizations have been added to the \gdsjoin algorithm since the first publication, including the use of Instruction-Level Parallelism (\textit{ILP}) in the distance calculation, which significantly improves the performance of the algorithm. Our experiments show that this newer version of \gdsjoin is more efficient than the published version~\cite{damon19}. Thus, we choose to use the newer more efficient version, as it is fairer than comparing \tensor with the original algorithm.


\section{Distance Calculations using Tensor Cores}
\label{sec:distance_calc_tensor}
We present our algorithm, \tensor, that leverages TCs for Euclidean distance calculations, and show how it is integrated into a distance similarity self-join algorithm. For illustrative purposes, in this section we use $4\times4$ matrices; however using the WMMA API and FP64, matrix sizes are either $8\times4$, $4\times8$ or $8\times8$.


    \subsection{Adapting the Euclidean Distance Formula}
    \label{sec:euclidean_tensor}
    Using the Euclidean distance formula defined above (Equation~\ref{eq:euclideanDistance}) between two points $a$ and $b$ in $d$ dimensions, we can expand this formula as follows:
    \begin{equation}
        dist(a, b) = \sqrt{(a_d - b_d)^2 + \ldots + (a_1 - b_1)^2 + 0}.
        \label{eq:euclidean_developed}
    \end{equation}
    We observe that, from right to left, the computation consists of a series of multiply-and-accumulate operations, where the distance in dimension $i$,
    computed as $(a_i - b_i)^2$ (hence a multiplication of two terms) gets accumulated with the distance previously computed in dimension $i - 1$, where $1 < i < d$. Let $a, b, c, d, e, f, g$ and $h$ be eight points in $d$ dimensions, and where we want to compute the Euclidean distance between $a, b, c, d$ and $e, f, g, h$. For illustration purposes only, we will use $4 \times 4$ matrices.
    \begin{figure}[h]
        \centering
        
        \begin{tikzpicture}[scale = 0.4]
    \draw (0, 0) rectangle (4,4);
    \node[anchor=west] at (0, 4.5) {$A$};
        \node[text=BrickRed] at (0.5, 3.5) {$a_1$};
        \node[text=BrickRed] at (1.5, 3.5) {$a_2$};
        \node[text=BrickRed] at (2.5, 3.5) {$a_3$};
        \node[text=BrickRed] at (3.5, 3.5) {$a_4$};
        
        \node[text=BrickRed] at (0.5, 2.5) {$a_1$};
        \node[text=BrickRed] at (1.5, 2.5) {$a_2$};
        \node[text=BrickRed] at (2.5, 2.5) {$a_3$};
        \node[text=BrickRed] at (3.5, 2.5) {$a_4$};
        
        \node[text=BrickRed] at (0.5, 1.5) {$a_1$};
        \node[text=BrickRed] at (1.5, 1.5) {$a_2$};
        \node[text=BrickRed] at (2.5, 1.5) {$a_3$};
        \node[text=BrickRed] at (3.5, 1.5) {$a_4$};
        
        \node[text=BrickRed] at (0.5, 0.5) {$a_1$};
        \node[text=BrickRed] at (1.5, 0.5) {$a_2$};
        \node[text=BrickRed] at (2.5, 0.5) {$a_3$};
        \node[text=BrickRed] at (3.5, 0.5) {$a_4$};
    
    \draw (5, 0)rectangle (9,4);
    \node[anchor=west] at (5, 4.5) {$B$};
        \node[text=RedOrange] at (5.5, 3.5) {$e_1$};
        \node[text=RedOrange] at (6.5, 3.5) {$e_2$};
        \node[text=RedOrange] at (7.5, 3.5) {$e_3$};
        \node[text=RedOrange] at (8.5, 3.5) {$e_4$};
        
        \node[text=green] at (5.5, 2.5) {$f_1$};
        \node[text=green] at (6.5, 2.5) {$f_2$};
        \node[text=green] at (7.5, 2.5) {$f_3$};
        \node[text=green] at (8.5, 2.5) {$f_4$};
        
        \node[text=cyan] at (5.5, 1.5) {$g_1$};
        \node[text=cyan] at (6.5, 1.5) {$g_2$};
        \node[text=cyan] at (7.5, 1.5) {$g_3$};
        \node[text=cyan] at (8.5, 1.5) {$g_4$};
        
        \node[text=magenta] at (5.5, 0.5) {$h_1$};
        \node[text=magenta] at (6.5, 0.5) {$h_2$};
        \node[text=magenta] at (7.5, 0.5) {$h_3$};
        \node[text=magenta] at (8.5, 0.5) {$h_4$};
        
    \node[anchor=west] at (9.5, 3) {\small{1. $B = B \times (-1.0)$ (CUDA cores)}};
    \node[anchor=west] at (9.5, 2) {\small{2. $C = A \times I + B$ (TCs)}};
    \node[anchor=west] at (9.5, 1) {\small{3. $D = C \times C^t + D$ (TCs)}};
\end{tikzpicture}
        \caption{Illustration of Euclidean distance calculations using TCs and Equation~\ref{eq:euclidean_developed}, between a point $a$ and four points $e, f, g, h$, and in four dimensions. This computation is computed in blocking fashion four dimensions at a time. Matrix $D$ contains the Euclidean distance between $a$ and the other points.}
        \label{fig:matrix_algo_1}
    \end{figure}
    
    We illustrate in Figure~\ref{fig:matrix_algo_1} how we can compute Euclidean distances using Equation~\ref{eq:euclideanDistance} and more particularly its equivalent, Equation~\ref{eq:euclidean_developed}, using TCs. Matrix $A$ contains a single point $a$ stored in row-major, while matrix $B$ can contain multiple points (here $e, f, g, h$), and is also row-major. To compute the difference between the coordinates, and to use TCs, we first scale $B$ by a factor $-1.0$, and we compute $C = A \times I + B$, where $I$ is the identity matrix. $C$ thus contains the difference between all coordinates of $a$ and the points $e, f, g$ and $h$, and in all four dimensions (because matrices are $4 \times 4$). We then multiply $C$ by its transpose, $C^t$, which computes the Euclidean distance between point $a$ and the points $e, f, g, h$, in the current dimensions that we store in $D$. This calculation is computed in blocking fashion four dimensions at a time.


    A severe limitation of using the Euclidean distance shown in Equation~\ref{eq:euclideanDistance} and represented in Figure~\ref{fig:matrix_algo_1}, is that it is only capable of computing the distance between one single point and several other points. Consider $D_{1,1}$ as the element in the first column of the first row of matrix $D$. The result of the computation in Figure~\ref{fig:matrix_algo_1} is that $D_{1,1} = dist(a, e)$, $D_{2,2} = dist(a, f)$, $D_{3,3} = dist(a, g)$ and $D_{4,4} = dist(a, h)$. Hence, out of the $4 \times 4 = 16$ results that matrix $D$ can store, only 4 correspond to actual Euclidean distances. Thus, while TCs have a higher peak throughput than CUDA cores~\cite{nvidia_a100}, only a fraction of the computation is actively used to compute Euclidean distances, which yields inefficient resource utilization. Furthermore, while we use $4 \times 4$ matrices for illustration purposes, we see in Figure~\ref{fig:matrix_algo_1} that all matrices used in the MMA operation need to have the same size, since the accumulator ($C$) is then used for the multiplication. However, when using the WMMA API and FP64, these matrix sizes are different and this solution can not be used. Consequently, we propose to use the expanded and equivalent form of the Euclidean distance outlined in Equation~\ref{eq:euclideanDistance}, which we detail as follows:

    \begin{equation}
        dist(a, b) = \sqrt{\sum^{d}_{i=1}a_{i}^2 - 2a_{i}b_{i} + b_{i}^2}.
        \label{eq:expanded_euclidean}
    \end{equation}
    Similarly to Equation~\ref{eq:euclidean_developed}, we can expand Equation~\ref{eq:expanded_euclidean}, yielding the following equation:
    \begin{equation}
        dist(a ,b) = \sqrt{\underbrace{a_{d}^2 + \overbrace{(-2a_{d}b_{d} + b_{d}^2)}^{\text{Tensor cores}}}_{\text{CUDA cores}} + \ldots + \underbrace{a_{1}^2 + \overbrace{(-2a_{1}b_{1} + b_{1}^2)}^{\text{Tensor cores}}}_{\text{CUDA cores}}}.
        \label{eq:euclidean_cuda_tensor}
    \end{equation}

    Using Equation~\ref{eq:euclidean_cuda_tensor}, we emphasize which part of the computation will be carried out by TCs and which part by the CUDA cores. Let $T_i = -2a_{i}b_{i} + b_{i}^2$ be the MMA operation done by TCs. To compute $dist(a_i, b_i)$, we need to calculate $a_i^2 + T_i$. To use TCs, we need to transform this into an MMA operation, computing either $a_i^2 \times I + T_i$, or $T_i \times I + a_i^2$, where $I$ is the identity matrix. However, as aforementioned, when using FP64 the WMMA API restricts us from reusing the accumulator from a previous MMA operation to be used in the multiplication of another MMA operation, due to different matrix sizes.
    Furthermore, using Equation~\ref{eq:euclidean_cuda_tensor}, we can compute the Euclidean distance between the four points $a, b, c, d$, and the four other points $e, f, g, h$ at a time, using the method illustrated in Figure~\ref{fig:matrix_algo_2}, and which was not possible using Equation~\ref{eq:euclideanDistance} (Figure~\ref{fig:matrix_algo_1}). 
    Finally, we observe that when computing the Euclidean distance between multiple points, and as will be the case when computing a distance similarity self-join for example, a part of the computation can be reused. The squared coordinates of the points ($a_i^2$ and $b_i^2$), are often reused throughout the computation. Indeed, the squared coordinates of a point are used for all the distance calculations with other points and do not change throughout the computation. Thus, the squared coordinates of the points can be precomputed to further improve the performance of the algorithm. As we still consider the use of $4 \times 4$ matrices for illustrative purposes, we store in an array $P$ the squared and accumulated coordinates of each point, four coordinates at a time. Considering that $a$ is the first point, the element 0 of this precomputed array $P$ is $a_1^2 + a_2^2 + a_3^2 + a_4^2$. For a dataset $V$ in $d$ dimensions, this array represents a memory overhead of only $|V| \times \lceil d / 4 \rceil$.


    \begin{figure}[h]
        \centering
        \begin{tikzpicture}[scale = 0.4]
    \draw (0, 0) rectangle (4,4);
    \node[anchor=west] at (0, 4.5) {$A$};
        \node[text=BrickRed] at (0.5, 3.5) {$a_1$};
        \node[text=BrickRed] at (1.5, 3.5) {$a_2$};
        \node[text=BrickRed] at (2.5, 3.5) {$a_3$};
        \node[text=BrickRed] at (3.5, 3.5) {$a_4$};
        
        \node[text=Green] at (0.5, 2.5) {$b_1$};
        \node[text=Green] at (1.5, 2.5) {$b_2$};
        \node[text=Green] at (2.5, 2.5) {$b_3$};
        \node[text=Green] at (3.5, 2.5) {$b_4$};
        
        \node[text=blue] at (0.5, 1.5) {$c_1$};
        \node[text=blue] at (1.5, 1.5) {$c_2$};
        \node[text=blue] at (2.5, 1.5) {$c_3$};
        \node[text=blue] at (3.5, 1.5) {$c_4$};
        
        \node[text=violet] at (0.5, 0.5) {$d_1$};
        \node[text=violet] at (1.5, 0.5) {$d_2$};
        \node[text=violet] at (2.5, 0.5) {$d_3$};
        \node[text=violet] at (3.5, 0.5) {$d_4$};
    
    \draw (5, 0)rectangle (9,4);
    \node[anchor=west] at (5, 4.5) {$B$};
        \node[text=RedOrange] at (5.5, 3.5) {$e_1$};
        \node[text=RedOrange] at (5.5, 2.5) {$e_2$};
        \node[text=RedOrange] at (5.5, 1.5) {$e_3$};
        \node[text=RedOrange] at (5.5, 0.5) {$e_4$};
        
        \node[text=green] at (6.5, 3.5) {$f_1$};
        \node[text=green] at (6.5, 2.5) {$f_2$};
        \node[text=green] at (6.5, 1.5) {$f_3$};
        \node[text=green] at (6.5, 0.5) {$f_4$};
        
        \node[text=cyan] at (7.5, 3.5) {$g_1$};
        \node[text=cyan] at (7.5, 2.5) {$g_2$};
        \node[text=cyan] at (7.5, 1.5) {$g_3$};
        \node[text=cyan] at (7.5, 0.5) {$g_4$};
        
        \node[text=magenta] at (8.5, 3.5) {$h_1$};
        \node[text=magenta] at (8.5, 2.5) {$h_2$};
        \node[text=magenta] at (8.5, 1.5) {$h_3$};
        \node[text=magenta] at (8.5, 0.5) {$h_4$};
    
    \draw (10, 0) rectangle (14,4);
    \node[anchor=west] at (10, 4.5) {$C$};
        \node[text=RedOrange] at (10.5, 3.5) {$e^2$};
        \node[text=RedOrange] at (10.5, 2.5) {$e^2$};
        \node[text=RedOrange] at (10.5, 1.5) {$e^2$};
        \node[text=RedOrange] at (10.5, 0.5) {$e^2$};
        
        \node[text=green] at (11.5, 3.5) {$f^2$};
        \node[text=green] at (11.5, 2.5) {$f^2$};
        \node[text=green] at (11.5, 1.5) {$f^2$};
        \node[text=green] at (11.5, 0.5) {$f^2$};
        
        \node[text=cyan] at (12.5, 3.5) {$g^2$};
        \node[text=cyan] at (12.5, 2.5) {$g^2$};
        \node[text=cyan] at (12.5, 1.5) {$g^2$};
        \node[text=cyan] at (12.5, 0.5) {$g^2$};
        
        \node[text=magenta] at (13.5, 3.5) {$h^2$};
        \node[text=magenta] at (13.5, 2.5) {$h^2$};
        \node[text=magenta] at (13.5, 1.5) {$h^2$};
        \node[text=magenta] at (13.5, 0.5) {$h^2$};
        
        
        
        
    
    \draw (15, 0) rectangle (19, 4);
    \node[anchor=west] at (15, 4.5) {$P'$};
        \node[text=BrickRed] at (15.5, 3.5) {$a^2$};
        \node[text=BrickRed] at (16.5, 3.5) {$a^2$};
        \node[text=BrickRed] at (17.5, 3.5) {$a^2$};
        \node[text=BrickRed] at (18.5, 3.5) {$a^2$};
        
        \node[text=Green] at (15.5, 2.5) {$b^2$};
        \node[text=Green] at (16.5, 2.5) {$b^2$};
        \node[text=Green] at (17.5, 2.5) {$b^2$};
        \node[text=Green] at (18.5, 2.5) {$b^2$};
        
        \node[text=blue] at (15.5, 1.5) {$c^2$};
        \node[text=blue] at (16.5, 1.5) {$c^2$};
        \node[text=blue] at (17.5, 1.5) {$c^2$};
        \node[text=blue] at (18.5, 1.5) {$c^2$};
        
        \node[text=violet] at (15.5, 0.5) {$d^2$};
        \node[text=violet] at (16.5, 0.5) {$d^2$};
        \node[text=violet] at (17.5, 0.5) {$d^2$};
        \node[text=violet] at (18.5, 0.5) {$d^2$};
        
    \node[anchor=north west] at (-0.25, -0.5) {\small{1. $A = A \times (-2.0)$ (CUDA cores)}};
    \node[anchor=north west] at (-0.25, -1.5) {\small{2. $T = A \times B + C$ (TCs)}};
    \node[anchor=north west] at (-0.25, -2.5) {\small{3. $D = D + T + P$ (CUDA cores)}};
\end{tikzpicture}
        \caption{Illustration of Euclidean distance calculations using TCs and Equation~\ref{eq:expanded_euclidean}, between four points $a, b, c, d$ and four points $e, f, g, h$, and in four dimensions. This computation is computed in blocking fashion four dimensions at a time. Matrix $D$ contains the Euclidean distances between $a, b, c, d$ and $e, f, g, h$.}
        \label{fig:matrix_algo_2}
    \end{figure}
    
    Figure~\ref{fig:matrix_algo_2} presents our algorithm design for computing the Euclidean distance between two sets of points, rather than between a single point and a set of points (Figure~\ref{fig:matrix_algo_1}). This method is based on Equation~\ref{eq:expanded_euclidean}. Matrix $A$ contains the first set of points ($a, b, c, d$), while $B$ contains the second set of points ($e, f, g, h$). Matrix $C$ contains the sum of squared coordinates of the points in $B$ and are pre-computed, as explained above. Matrix $P'$ contains the sum of squared coordinates of the points in $A$. Our algorithm first scales matrix $A$, and then computes $T = A \times B + C$ using TCs. We then use the CUDA cores to accumulate $P'$, as well as the result matrix $D$. Because $C, D$, and $T$ are different sizes than $A$ and $B$, we can not use TCs to compute these operations, which is a limitation of the WMMA API when using FP64. This computation is computed in blocking fashion four dimensions at a time. The algorithm outputs matrix $D$ which contains the Euclidean distance between $a, b, c, d$ and $e, f, g, h$, which corresponds to 16 distances, compared to only 4 when using the algorithm shown in Figure~\ref{fig:matrix_algo_1}. While we illustrate the computation using $4 \times 4$ matrices, when using the WMMA API, because $D$ is an $8 \times 8$ matrix, we can compute 64 distances instead of 8.

    \subsection{Tensor Cores for Distance Similarity Joins}
    \label{sec:tensor_join}
    As we outlined in Section~\ref{sec:distance_simjoin}, most of the distance similarity self-join algorithms in the literature reduce the overall computational complexity by using an index data structure and, compared to a brute-force approach, typically reduces the number of candidate points that need to be refined per query point. In particular, the distance similarity self-join algorithm that we leverage here, \gdsjoin, uses a grid index with cells of size $\epsilon^d$. For each query point in the dataset $V$, we thus \textit{search} the grid indexing for neighboring cells, yielding a set of candidate points for each of the query points, which are then \textit{refined} by computing the Euclidean distance between them and the query point. Because \tensor and \gdsjoin use the same index, both algorithms yield the same candidate points to  be refined using distance calculations. This allows us to compare the performance of CUDA and TCs in a self-consistent manner, where the performance differences are directly attributable to distance calculations. 
    
    A characteristic of the grid index we are using is that all the query points from the same cell share the same candidate points. This characteristic is particularly important, as it is necessary to efficiently make use of Equation~\ref{eq:euclidean_cuda_tensor} (Figure~\ref{fig:matrix_algo_2}). Indeed, the query points we use in matrix $A$ must compute their Euclidean distances, in matrix $B$, to the same set of candidate points. Hence, the query points used in matrix $A$ should come from the same grid cell, as they share the same set of candidate points.
    
    Another optimization used by Gowanlock and Karsin~\cite{damon19} is the batching of the execution. Because the final result of the similarity self-join might exceed the memory size of the GPU, the entire execution is split across multiple batches. As a positive side-effect, multiple batches allow for hiding data transfers between the host and the GPU with computation. Indeed, batches are computed by several parallel CUDA streams, where the data transfers of a stream can overlap the computation of another stream. However, as a batch corresponds to a set of query points to compute, we must ensure in our case that the query points we send in a batch can be computed by our TCs algorithm. More specifically, when assigning query points from a batch to a warp on the GPU, we must ensure that these query points belong in the same grid cell and are not from different cells. Otherwise, we would be unable to use the algorithm presented in Figure~\ref{fig:matrix_algo_2}.
    
    Using the WMMA API and FP64, only one combination of matrix sizes is available. Namely, matrix $A$ will contain up to four coordinates of up to eight query points, matrix $B$ up to four coordinates of up to eight candidate points, matrix $C$ the sum of squared coordinates of up to eight candidate points, and matrix $D$ up to sixty-four Euclidean distances. Because TCs operate at a warp level using the WMMA API, we assign up to eight query points to a warp, which will then compute the Euclidean distance to all the candidate points, as determined by the use of the grid index. If the number of query points, candidate points, or coordinates is insufficient to fill the remaining rows or columns of the matrices, we must fill them with zeros. Because we process four dimensions at a time, up to $\lceil d / 4 \rceil$ steps are necessary to compute the Euclidean distance. Similarly to \gdsjoin~\cite{damon19}, we enable distance calculations short-circuiting, which may happen after every MMA operation, i.e., for every 4 dimensions. However, all currently computed Euclidean distances between all the query points and candidate points of the warp must short-circuit to trigger this optimization.


\section{Experimental Evaluation}
\label{sec:exp_eval}
In this section, we detail the experimental evaluation we conducted. We start by comparing our TCs algorithm and another optimized TCs algorithm to compute Euclidean distances. We then compare our proposed algorithm \tensor to other state-of-the-art distance similarity self-join algorithms.


    \subsection{Datasets}
    \label{sec:datasets}
    We evaluate the algorithms using a wide range of real-world and synthetic datasets, spanning several sizes, dimensionalities, and distributions. Synthetic datasets are generated following either a uniform or exponential distribution, and their name is prefixed by either \textit{Unif} or \textit{Expo}, respectively, followed by the dimensionality and the number of points (e.g., \textit{Expo3D2M} is an exponentially distributed 3-D dataset containing 2M points). We summarize the different synthetic datasets that we use in Table~\ref{tab:datasets_synth}, and the real-world datasets in Table~\ref{tab:datasets_real}. \textit{Gaia50M} and \textit{OSM50M} are the first 50M points of the original datasets, as described by Gowanlock~\cite{bps_clustering}. We choose to use different distributions to better evaluate the performance of TCs under different workloads: when a dataset is uniformly distributed, TCs should all have a similar workload, while when a dataset is exponentially distributed, some TCs will have a higher workload than other TCs.
    
    \begin{table}
        \centering
        \caption{Synthetic datasets used in the experimental evaluation.}
        \begin{tabularx}{\columnwidth}{Xll} \hline
            Distribution & $d$ & $n$ \\ \hline
            Uniform & 2, 3, 4, 6, 8 & 10M \\
            Exponential & 2, 3, 4, 6, 8 & 2M, 10M \\ \hline
        \end{tabularx}
        \label{tab:datasets_synth}
    \end{table}

    \begin{table}
        \centering
        \caption{Real-world datasets used in the experimental evaluation.}
        \begin{tabularx}{\columnwidth}{XllXll} \hline
            Dataset & $d$ & $n$ & Dataset & $d$ & $n$  \\ \hline
            \textit{SW2DA}~\cite{spaceWeather} & 2 & 1.86M & \textit{SW2DB}~\cite{spaceWeather} & 2 & 5.16M \\
            \textit{OSM50M}~\cite{osm} & 2 & 50M & \textit{Gaia50M}~\cite{gaia} & 2 & 50M \\
            \textit{SW3DA}~\cite{spaceWeather} & 3 & 1.86M & \textit{SuSy}~\cite{susy} & 18 & 5M \\
            \textit{BigCross}~\cite{bigcross} & 57 & 11M & \textit{Songs}~\cite{songs} & 90 & 515K \\ \hline
        \end{tabularx}
        \label{tab:datasets_real}
    \end{table}
    
    We denote the selectivity as $s$, which represents the average number of neighboring points found within $\epsilon$ of each query point when performing a similarity self-join, excluding each query point finding itself. The selectivity is calculated as follows: $s = (|R| - |V|) / |V|$, where $|R|$ and $|V|$ are the result set of the similarity self-join and dataset sizes, respectively. This metric is used in the literature to quantify the complexity of the search for a given value of $\epsilon$: increasing $\epsilon$ results in more work to compute, and a higher selectivity. 


    \subsection{Methodology}
    \label{sec:methodology}
    We conducted our experiments on the following platforms: \textbf{Platform 1}: $2\times$ AMD Epyc 7542 CPU ($2 \times 32$ cores, 2.9GHz), 512~GiB of RAM, Nvidia A100 GPU; \textbf{Platform 2:} Intel Xeon W-2295 CPU (18 cores, 3GHz), 256~GiB of RAM.
        
    In this section, we use the distance similarity join application as a case study for the use of TCs for Euclidean distance calculations. For completeness, we compare our algorithm to other distance similarity join algorithms, including parallel CPU algorithms. However, this is only one example application, and thus we also show brute-force CUDA vs. TC performance as it may be more applicable to other algorithms.

    The algorithms \tensor, \gdsjoin, \superego and \hilbert are configured as follows:
    
        \noindent\textbullet \tensor: Our proposed TCs algorithm is executed on Platform~1, configured with 256 threads per block (8 warps), up to 8 query points per warp, and using distance calculations short-circuiting, as explained in Section~\ref{sec:tensor_join}.
        
        \noindent\textbullet \gdsjoin: Parallel GPU algorithm proposed by Gowanlock and Karsin~\cite{damon19} and further optimized since the original publication, executed in Platform~1. This algorithm is configured with 256 threads per block, $ILP = min(8,d)$ and uses distance calculations short-circuiting, as presented in Section~\ref{sec:distance_simjoin}.
        
        \noindent\textbullet \superego: Parallel CPU algorithm proposed by Kalashnikov~\cite{superego}, optimized by Gallet and Gowanlock~\cite{hegjoin} and executed on Platform~1 using 64 threads (the number of physical cores on the platform).
        
        \noindent\textbullet \hilbert: Parallel CPU algorithm proposed by Perdacher et al.~\cite{fgfHilbert}, executed on Platform~2 (the only platform supporting AVX-512, required for this algorithm) and using 18 threads (the number of physical cores on the platform).
        
    While we would have preferred to use a single platform to conduct all our experiments, and thus have the same number of threads/cores for all CPU algorithms, prior experiments we conducted showed us that both \superego and \hilbert had a relatively poor scalability. Hence, if we were able to run \hilbert using 64 threads/cores, as we did for \superego, the results we show in the following sections would not have been significantly different. Furthermore, note that despite using fewer threads/cores, \hilbert typically outperforms \superego.
    
    All the algorithms are using double precision (FP64) to compute, and are compiled using NVCC v11.2 (for \tensor and \gdsjoin) or GCC (v8.5 for \superego, and v9.4 \hilbert) using the O3 optimization.
    
    During our experiments, many scenarios using \hilbert did not produce the correct self-join results, which are consequently not included. We believe that the issues encountered with \hilbert are due to the width of the vectorized instructions: 512-bits, or 8 FP64 values, which may not be working when $d < 8$. Furthermore, \superego happened to fail in several low-dimensional cases without a clear understanding of the reason, and we thus also do not report the execution time of these experiments. However, we consider that the successful experiments should be sufficient to accurately evaluate the performance of \tensor compared to the other algorithms. Finally, note that the four algorithms, \tensor\footnote{\url{https://github.com/benoitgallet/ted-join-hipc22}}, \gdsjoin~\cite{damon19}, \superego~\cite{superego}, and \hilbert~\cite{fgfHilbert}, are publicly available.


    
    \subsection{Results: Comparison of Brute-force TC Approaches}
    \label{sec:results_bf}
    
    We compare the performance of TCs and CUDA cores for performing Euclidean distance calculations when using brute-force computation, which is $O(|V|^2)$. Here, we use the algorithm \tensor (TCs), to which we removed all optimizations, including indexing, and compare it to a highly optimized MMA reference implementation by Nvidia~\cite{wmmaNvidia}, that leverages the WMMA API similarly to \tensor, denoted as \wmmaRef. We selected this implementation instead of a library such as cuBLAS~\footnote{\url{https://developer.nvidia.com/cublas}} or CUTLASS~\footnote{\url{https://github.com/NVIDIA/cutlass}} (with the latter built upon the WMMA API), as it is the best direct comparison between approaches.

    We outline two major differences between \wmmaRef and \tensor, as a consequence of matrix size, as follows:
    \begin{enumerate}[leftmargin=*]
        \item The matrix sizes are dependent on data dimensionality and impact performance~\cite{tensorPerfPrecisionGTC, tensorPerfUltimateGuideGTC}. \wmmaRef is designed and optimized for large MMA operations, whereas \tensor targets smaller matrices. 
        \item \tensor uses small matrices, and thus computes many small $8 \times 8$ distance matrices and leverages shared memory. In contrast, \wmmaRef computes the entire $|D|^2$ distance matrix, thus requiring a much larger memory footprint. Consequently, when using \wmmaRef, and to be able to use it on large datasets that would exceed global memory capacity, we store the result matrix using unified memory, which automatically pages data between main and global memory. Furthermore, as cuBLAS and CUTLASS work similarly to \wmmaRef, they have the same drawback related to the use of unified memory.
    \end{enumerate}
    
\begin{figure}[!t]
        \centering
        \includegraphics[scale = 0.29]{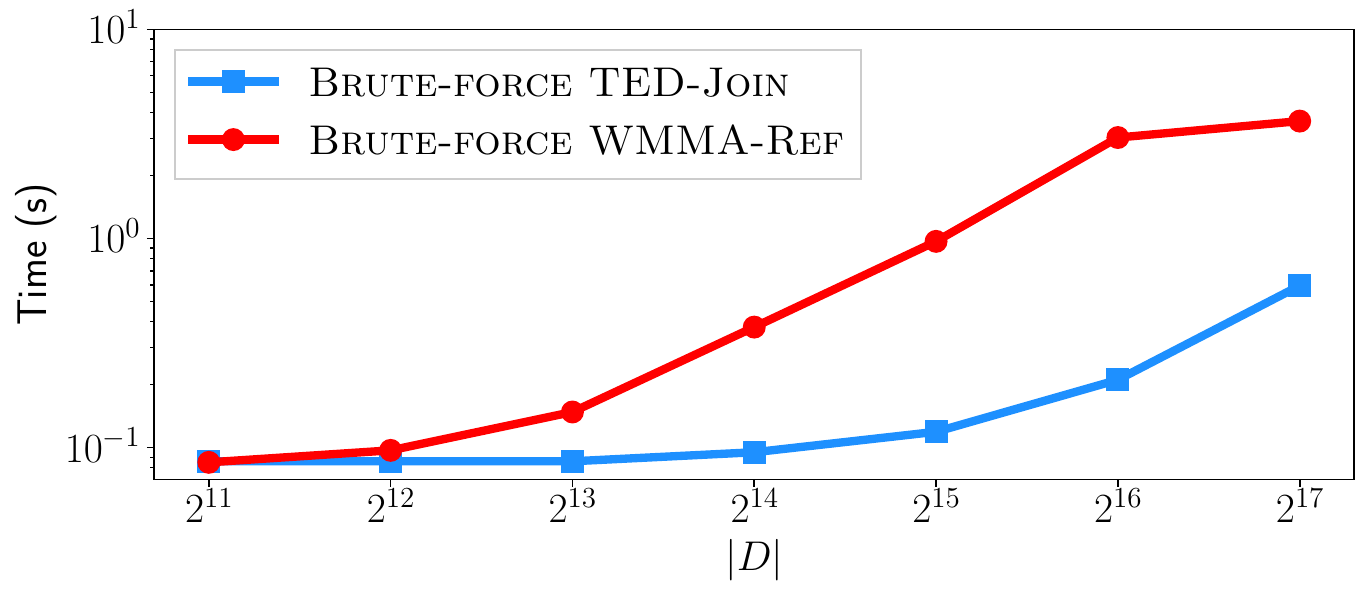}
        \caption{Response time of our proposed algorithm \tensor, and \wmmaRef an optimized MMA algorithm from Nvidia leveraging the WMMA API, using brute-force searches to compute Euclidean distance calculations on a 16-D exponentially distributed synthetic datasets.}
        \label{fig:brute_force}
    \end{figure}

    Figure~\ref{fig:brute_force} plots the performance of \tensor and \wmmaRef using brute-force searches (i.e., without using an index) to compute Euclidean distance calculations on a 16-D exponentially distributed synthetic datasets, spanning $2^{11}$ to $2^{17}$ points (we omit datasets with other dimensionalities as we observed similar results). Note that $2^{18}$ points overflows main memory when using \wmmaRef. We observe that the performance of \wmmaRef degrades quicker than \tensor as the dataset size increases. We attribute these results to the use of unified memory by \wmmaRef, which is required to store the large result matrix ($|D|\times|D|$), and which is paged between GPU global and main memory when its size exceeds global memory capacity. In addition to the poor performance attributed to unified memory, using \wmmaRef, which computes on large matrices and thus on the $d$ dimensions of a dataset at a time, limits the use of several optimizations, which are explored in the following sections. Namely, this inhibits short-circuiting the distance calculations when the cumulative distance between points exceeds $\epsilon$.
    
    We profile \tensor and \wmmaRef on the $2^{17}$ points 16-D dataset (Figure~\ref{fig:brute_force}). With this dataset size, unified memory needs to be paged between global and main memory throughout the execution. We measure that \wmmaRef transfers 687.84~GB between the L1 and L2 caches, and 503.61~GB between the L2 cache and global memory. In comparison \tensor transfers 558.57~GB and only 0.046~GB, respectively, as we rely on shared memory to store small ($8\times8$) result matrices, rather than a large $|V|\times|V|$ matrix in global memory like \wmmaRef. This results in lower L1 and L2 hit rates: $19.35\%$ using \wmmaRef vs. $50.32\%$ using \tensor for the L1 hit rate, and $72.45\%$ vs. $99.99\%$ for the L2 hit rate. In summary, the unified memory required by \wmmaRef negatively affects performance in the case of distance calculations, and thus \tensor should be preferred.
    
    \subsection{Results: Optimized TC and CUDA Core Approaches}
    \label{sec:results2}
    We investigate in this section the performance of \tensor, as compared to other state-of-the-art algorithms from the literature: \gdsjoin, \superego, and \hilbert.
    
    \subsubsection{Uniformly Distributed Datasets}
    \label{sec:results_uniform}
    We start this result section with uniformly distributed synthetic datasets, detailed in Table~\ref{tab:datasets_synth}. We select this distribution as all the query points will have a similar number of candidate points to refine, allowing us to evaluate the performance of TCs when their workload is relatively uniform.
    
\begin{figure}[!t]
        \centering
        \includegraphics[scale = 0.35]{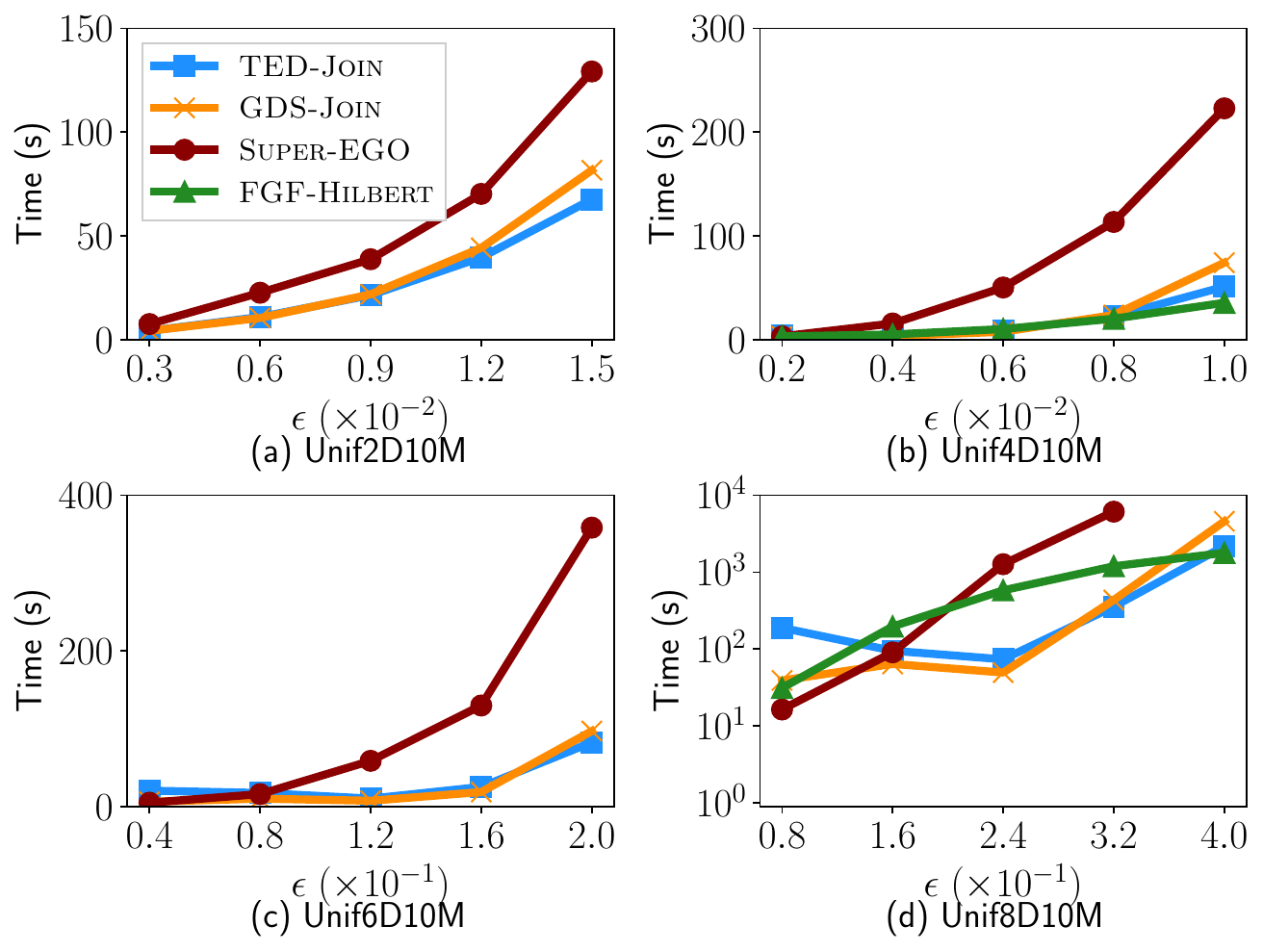}
        \caption{Response times of the \tensor, \gdsjoin, \superego, and \hilbert on a selection of uniformly distributed synthetic datasets. $s$ is in the range (a) $282$--$6978$, (b) $71$--$8449$, (c) $7$--$4295$ and (d) $0$--$10888$. The legend in (a) corresponds to all subfigures. $d \in \{2,4,6,8\}$, $n=10$M.}
        \label{fig:unif_full}
    \end{figure}

    We show in Figure~\ref{fig:unif_full} the execution time of \tensor compared to \gdsjoin, \superego, and \hilbert on a selection of uniformly distributed synthetic datasets. In these cases, we can see that \superego is consistently performing worse than all of the other algorithms, except on the \textit{Unif8D10M} dataset when $\epsilon = 0.08$ (Figure~\ref{fig:unif_full}(d)). Furthermore, we observe that \tensor performs similarly or better than \gdsjoin in most cases, except on \textit{Unif8D10M} when $\epsilon < 0.32$. From these results, it seems that \tensor performs similar to \gdsjoin when $\epsilon$ is low, and therefore when the workload is low as well, potentially indicating an overhead from using TCs. But when $\epsilon$ increases, and thus the workload, the higher computational throughput of TCs outperforms the CUDA cores used by \gdsjoin.
    
    We also observe that the speedup is the highest on the 2-D and 4-D datasets since all 2 or 4 dimensions can be computed at once using TCs, as we compute 4 dimensions at a time. The speedup is the lowest on the 6-D datasets since we need to compute the distances in two iterations (as many as for the 8-D datasets), but where 2 dimensions are zeros and thus that the CUDA cores in \gdsjoin do not have to compute.
    
    \subsubsection{Exponentially Distributed Datasets}
    \label{sec:results_expo}
    In this section we present the results on the same algorithms as in Section~\ref{sec:results_uniform} on the exponentially distributed synthetic datasets, detailed in Table~\ref{tab:datasets_synth}. We select this distribution as it creates a large workload variance between the query points, where some query points may have many candidate points to refine, and other query points very few, which allows us to evaluate the performance of the TCs when their workload varies. 
    
\begin{figure}[!t]
        \centering
        \includegraphics[scale = 0.35]{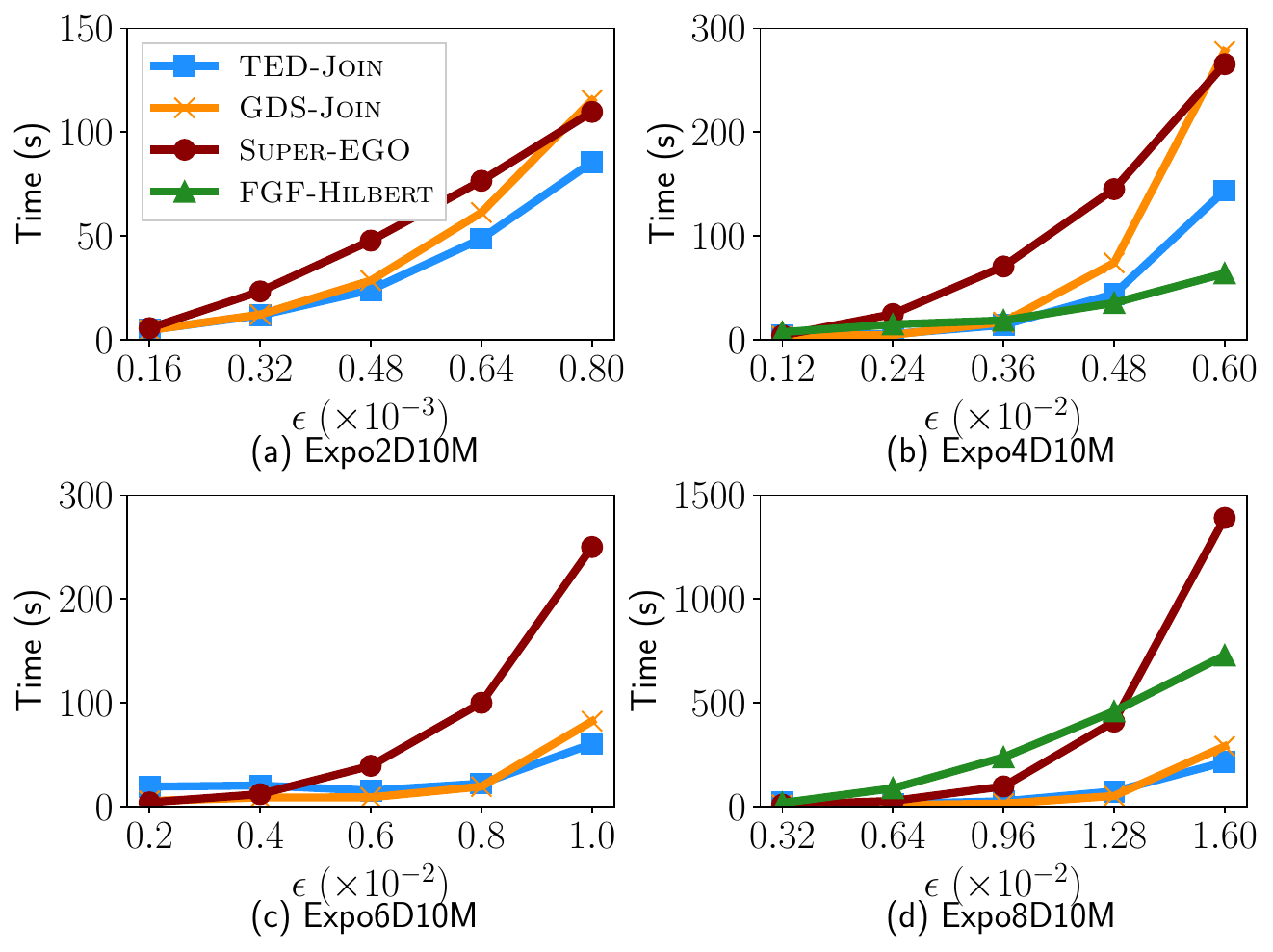}
        \caption{Response times of \tensor, \gdsjoin, \superego, and \hilbert on a selection of exponentially distributed synthetic datasets. $s$ is in the range (a) $320$--$7834$, (b) $15$--$7414$, (c) $0$--$1658$ and (d) $0$--$1210$. The legend in (a) corresponds to all subfigures. $d \in \{2,4,6,8\}$, $n=10$M.}
        \label{fig:expo_full}
    \end{figure}

    Figure~\ref{fig:expo_full} reports the execution time of \tensor compared to \gdsjoin, \superego, and \hilbert on a selection of exponentially distributed synthetic datasets. Note that \hilbert did not run correctly on the 2-D and 6-D datasets (Figures~\ref{fig:expo_full}(a) and~(c)). In these experiments, \tensor typically performs similarly or better than \gdsjoin, particularly as $\epsilon$ increases. \superego is consistently outperformed by the other algorithms, while \hilbert performs the best on the \textit{Expo4D10M} dataset (Figure~\ref{fig:expo_full}(b)), but is outperformed by both \tensor and \gdsjoin on the \textit{Expo8D10M} dataset (Figure~\ref{fig:expo_full}(d)). Because these datasets are exponentially distributed, the workload throughout the computation of the similarity self-join can vary a lot. The query points in the denser regions of the dataset will have many candidate points to refine, and the query points in the sparse regions of the dataset may have only a few candidate points. Hence, and despite a highly varying workload, \tensor remains more efficient in most cases compared to \gdsjoin and all compared algorithms in general, particularly in lower dimensions ($2 \leq d \leq 4$).
    
    \subsubsection{Real-World Datasets}
    \label{sec:results_real}
\begin{figure}[!t]
        \centering
        \includegraphics[scale = 0.35]{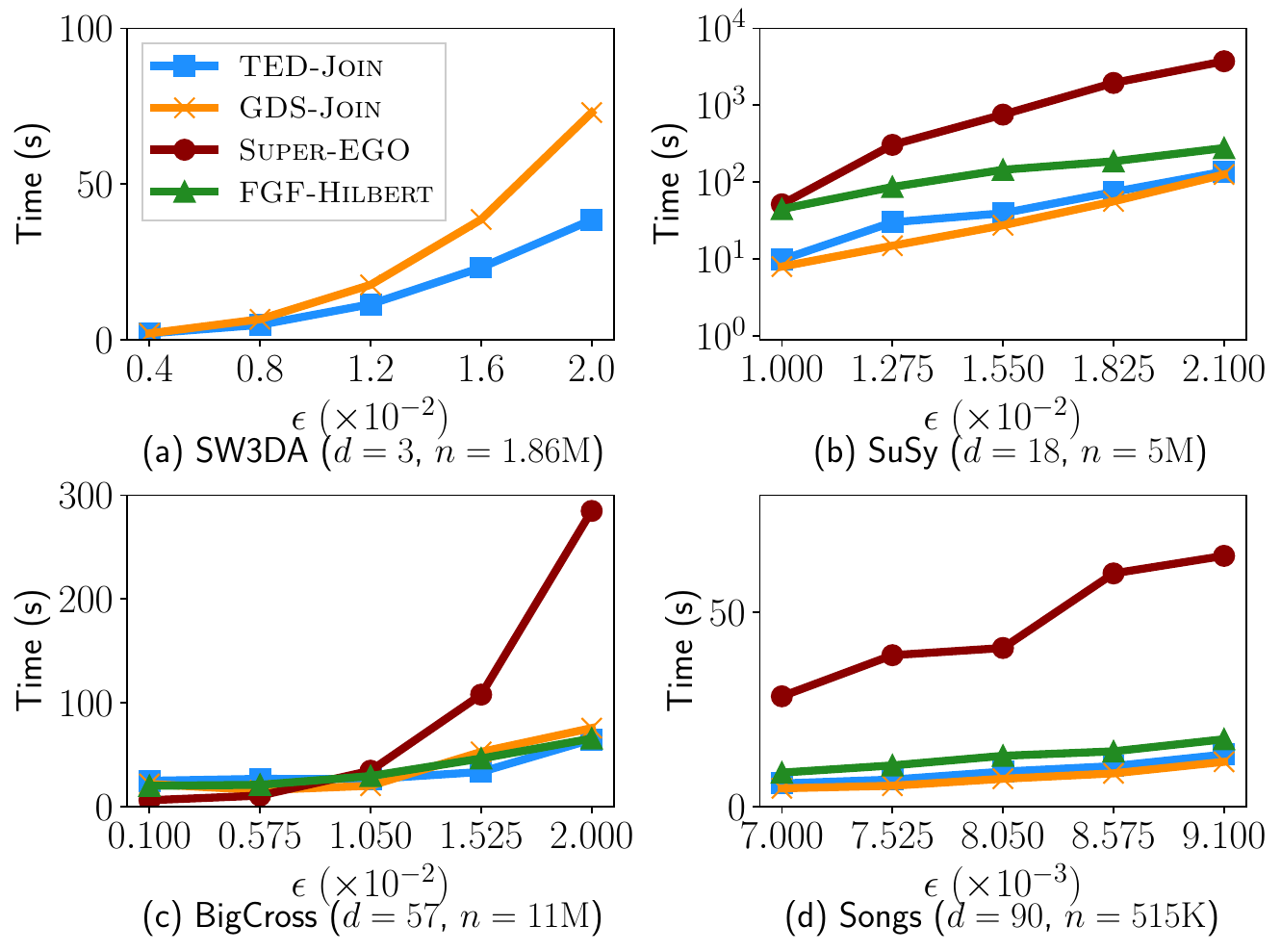}
        \caption{Response times of \tensor, \gdsjoin, \superego, and \hilbert on a selection of real-world datasets (Table~\ref{tab:datasets_real}). $s$ is in the range (a) $163$--$5373$, (b) $5$--$1090$, (c) $1$--$1104$ and (d) $127$--$998$. The legend in (a) corresponds to all subfigures.}
        \label{fig:real_full}
    \end{figure}

    We present in this section the results of \tensor, \gdsjoin, \superego and \hilbert on a selection of the real-world datasets (Table~\ref{tab:datasets_real}), as shown in Figure~\ref{fig:real_full}. \tensor and \gdsjoin perform very similarly, particularly on the higher dimensional datasets (Figures~\ref{fig:real_full}(b)--(d)), while \tensor outperforms \gdsjoin on the \textit{SW3DA} dataset as $\epsilon$ increases (Figure~\ref{fig:real_full}(a)). \hilbert also performs quite similarly to \tensor and \gdsjoin, while \superego is often outperformed by the other algorithms. Overall, these experiments show that \tensor and \gdsjoin may perform similarly as dimensionality increases, while \tensor yields an advantage in lower dimensions (Figure~\ref{fig:real_full}(a)), as we observed in previous Figures~\ref{fig:unif_full} and~\ref{fig:expo_full}. These experiments show us that in higher dimensions (Figures~\ref{fig:real_full}(b)--(d)), \tensor may not yield an advantage compared to \gdsjoin .
    
    \subsection{Discussion: When Tensor Cores Should Be Employed}
    \label{sec:results_speedup}

    \begin{figure}[!t]
        \centering
        \includegraphics[scale=0.375]{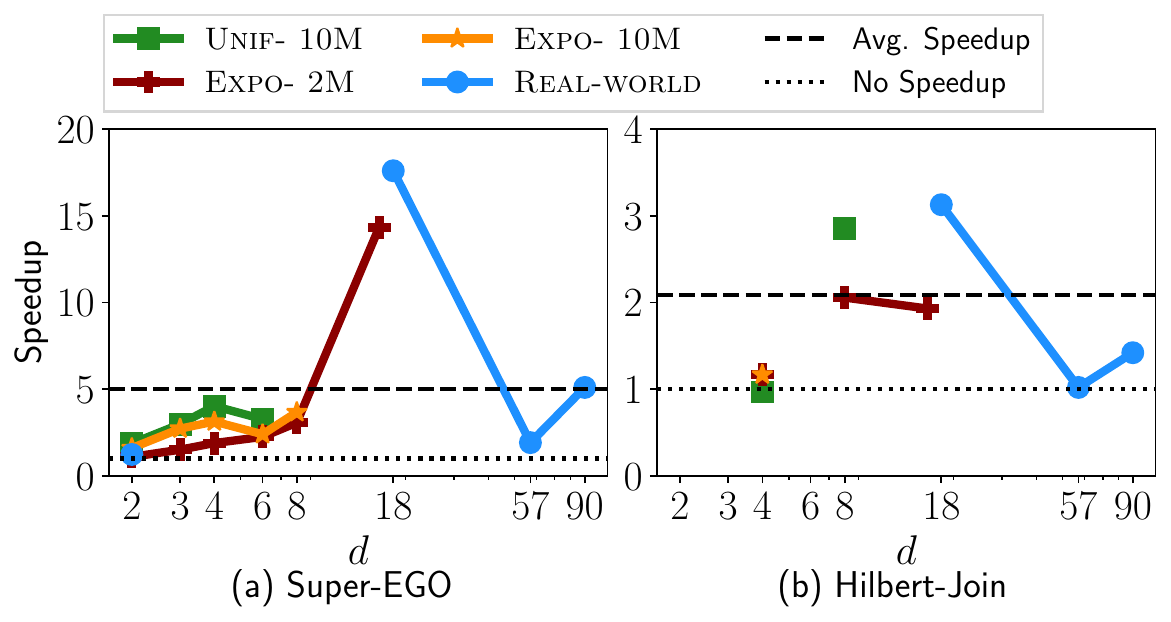}
        \caption{Speedups of \tensor over (a)~\superego and (b)~\hilbert across datasets presented in Tables~\ref{tab:datasets_synth} and~\ref{tab:datasets_real}, for all values of $\epsilon$ we used, and as a function of the dimensionality. The dashed horizontal lines correspond to the average speedups of \tensor over a compared algorithm, and the dotted horizontal lines represent no speedup.}
        \label{fig:speedup_cpu}
    \end{figure}
    
    \begin{figure}[!t]
        \centering
        \includegraphics[scale=0.325]{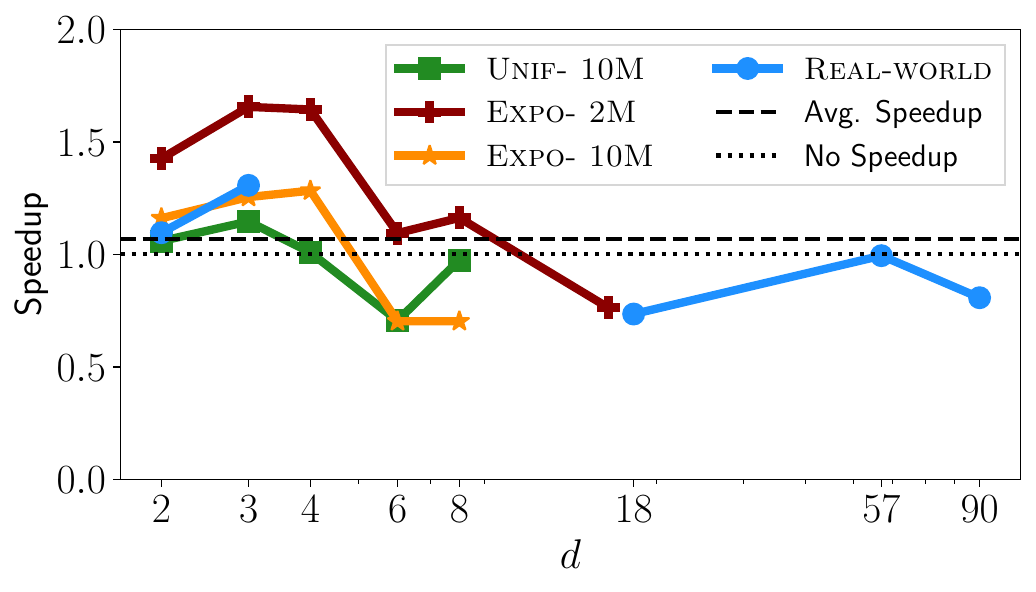}
        \caption{The same as for Figure~\ref{fig:speedup_cpu}, but plotting the speedup of \tensor over \gdsjoin.}
        \label{fig:speedup_gds}
    \end{figure}
    
    We summarize the results of \tensor as compared to the \superego~\cite{hegjoin}, \hilbert~\cite{fgfHilbert}, and \gdsjoin~\cite{damon19} algorithms that we obtained across experiments, including those that were omitted due to space constraints. The experiments covered a wide range of data dimensionalities, sizes, and distributions, resulting in an insightful picture of the overall performance of using TCs in \tensor compared to the use of CUDA cores in \gdsjoin. We report the speedup of \tensor over the \superego, \hilbert, and \gdsjoin algorithms in Figures~\ref{fig:speedup_cpu}(a) and~(b), and Figure~\ref{fig:speedup_gds}, respectively. We also report the L1 and L2 cache hit rates of \gdsjoin and \tensor in Table~\ref{tab:profiling}, and the average and maximum speedups of \tensor over \superego, \hilbert, and \gdsjoin in Table~\ref{tab:speedup}.
    
    \begin{table}[!t]
        \centering
        \caption{L1 and L2 cache hit rates of \gdsjoin and \tensor on a selection of exponentially distributed synthetic datasets ($2 \leq d \leq 16$, $n=2$M) and real-world datasets (SW3DA and SuSy), measured using the Nvidia Nsight Compute profiler.}
        \begin{tabularx}{\columnwidth}{XXXXX}
            \cline{2-5}
                    & \multicolumn{2}{c}{\gdsjoin} & \multicolumn{2}{c}{\tensor} \\ \hline
            \multicolumn{1}{X}{\textit{Dataset}} & L1 & L2 & L1 & L2  \\ \hline
            \multicolumn{1}{X}{\textit{Expo2D2M}} & 71.55\% & 97.85\% & 66.40\% & 98.11\% \\
            \multicolumn{1}{X}{\textit{Expo4D2M}} & 89.89\% & 95.60\% & 67.20\% & 97.65\% \\
            \multicolumn{1}{X}{\textit{Expo8D2M}} & 90.53\% & 97.15\% & 45.76\% & 66.23\% \\
            \multicolumn{1}{X}{\textit{Expo16D2M}} & 97.27\% & 99.84\% & 52.15\% & 57.27\% \\
            \multicolumn{1}{X}{\textit{SW3DA}}   & 68.84\% & 97.62\% & 54.70\% & 94.43\% \\
            \multicolumn{1}{X}{\textit{SuSy}}    & 92.13\% & 86.91\% & 38.00\% & 53.50\% \\ \hline
        \end{tabularx}
        \label{tab:profiling}
    \end{table}

    \begin{table}[!t]
        \centering
        \caption{Average and maximum speedup of \tensor over \superego, \hilbert, and \gdsjoin across experiments reported in Figures~\ref{fig:speedup_cpu} and~\ref{fig:speedup_gds}.}
        \begin{tabularx}{\columnwidth}{Xlll}
            \cline{2-4}
                                        & \multicolumn{2}{c}{CPU} & \multicolumn{1}{c}{GPU} \\
            \cline{2-4}
                                        & \superego         & \hilbert          & \gdsjoin ($d \leq 4$)         \\ \hline
            \multicolumn{1}{X}{Average} & $5.00\times$      & $2.09\times$      & $1.07\times$ ($1.28\times$)   \\
            \multicolumn{1}{X}{Maximum} & $27.22\times$     & $9.46\times$      & $2.23\times$ ($2.23\times$)   \\ \hline
        \end{tabularx}
        \label{tab:speedup}
    \end{table}
    
    Figure~\ref{fig:speedup_cpu}(a) plots the speedup of \tensor over the CPU algorithm \superego~\cite{superego, hegjoin}. We observe that \tensor consistently achieves a speedup $>1$, with an average of $5.00\times$ and a maximum of $27.22\times$. Thus, we believe that there is no clear disadvantage to using \tensor over \superego, regardless of the dimensionality, dataset distribution, or size.
    
    Figure~\ref{fig:speedup_cpu}(b) plots the speedup of \tensor over the CPU algorithm \hilbert~\cite{fgfHilbert}. Because many of our experiments could not be correctly conducted using the \hilbert algorithm, it makes it harder to draw a clear conclusion regarding the performance \tensor compared to \hilbert. However, in the successful experiments, our TCs solution achieved an average speedup of $2.09\times$ with a maximum of $9.46\times$, and the majority of the speedups are above 1. Hence, and similarly to \superego, there is no clear disadvantage of using \tensor over \hilbert.
    
    Observing the speedup of \tensor over the CUDA core algorithm \gdsjoin (Figure~\ref{fig:speedup_gds}), we achieve the best performance when $d \leq 4$, and is best on exponentially distributed synthetic and real-world datasets. However, as the dimensionality $d$ increases, this speedup decreases, resulting in an average speedup of only $1.07\times$, but achieving a maximum of $2.23\times$ on the \textit{Expo3D2M} dataset. If we only consider datasets where $d \leq 4$, \tensor achieves an average speedup of $1.28\times$ over \gdsjoin. 
    Regarding the relatively low speedup in higher dimensions, TCs are designed for large matrix multiplications, where data can be reused when computing tiles of the resulting matrix. In the case of \tensor, we are unable to reuse such data, thus limiting the performance. Additionally, we measure and compare the L1 and L2 cache hit rates of \gdsjoin and \tensor (Table~\ref{tab:profiling}). While \gdsjoin consistently achieves high cache hit rates, as the dimensionality increases, the cache hit rate using \tensor decreases significantly. This explains why the speedup of \tensor over \gdsjoin decreases with increasing dimensionality (Figure~\ref{fig:speedup_gds}).
    
    From these results, we conclude that TCs should be used when the dimensionality is low ($2 \leq d \leq 4$). Furthermore, there are cases where the dimensionality does not evenly divide by 4 (the dimension of the matrices as defined by the WMMA API for FP64). In total, $\lceil d / 4 \rceil$ MMA operations are needed to compute distance calculations, meaning that an additional MMA operation needs to be performed for cases where $d$ mod $4 \neq 0$, which performs excess work. For example, because 6-D datasets are stored as 8-D datasets, where the last two dimensions are filled with zeros, TCs cannot achieve peak performance.
    
    In summary, TCs should be used under the following scenarios instead of the reference implementations on their respective architectures:
    
    \noindent\textbullet Compared to using CUDA cores, TCs should be used on low-dimensional datasets ($2 \leq d \leq 4$).
    
    \noindent\textbullet There is no drawback of using TCs over multi-core CPUs.


\section{Conclusion and Future Work}
\label{sec:conclusion}
In this paper, we presented a novel approach to computing Euclidean distances leveraging TCs on Nvidia GPUs. TCs are designed solely for Matrix Multiply-Accumulate operations, and yield a much higher peak throughput than CUDA cores for this operation~\cite{nvidia_a100}. While TCs have been extensively used in fields such as machine learning, their usage remains very limited for more general-purpose applications. Hence, to our knowledge, this paper presents the first use of TCs for FP64 Euclidean distance calculations, where FP64 TCs computation has only been possible using the Ampere generation of Nvidia GPUs. This makes our algorithm suitable for scenarios where precise computation using FP64 is required. As such, our algorithm can provide the foundation for improving the performance of other data analysis applications where distance calculations are used (e.g., distance similarity searches, $k$NN, and \dbscan~\cite{damon19, superego, fgfHilbert, hegjoin, gowanlock_knn, dbscan}). In these cases, our TC GPU kernel can be adapted to refine candidate points independently of the index that is used.

\textbf{Comparison to tensor algorithms:} we compared \tensor to a reference MMA implementation, \wmmaRef, from Nvidia~\cite{wmmaNvidia}, where no optimizations (including an index) were used. We find that \tensor outperforms \wmmaRef, because the latter requires unified memory to store a $|V|\times|V|$ distance matrix. Libraries such as cuBLAS and CUTLASS have the same drawbacks as \wmmaRef, and are thus also unsuitable for moderately sized input datasets.

\textbf{Comparison to similarity search reference implementations:} we compared \tensor to the GPU algorithm \gdsjoin~\cite{damon19}. Despite an average speedup of $1.07\times$ over \gdsjoin when $2 \leq d \leq 90$, we achieve a maximum speedup of $2.23\times$ over this algorithm. We find that \tensor yields the best performance when $d \leq 4$ with an average speedup of $1.28\times$ over \gdsjoin. Because \tensor and \gdsjoin use the same index, this performance improvement is a direct result of employing TCs. While the maximum speedup is expected to be $2\times$ due to the maximum throughput of TCs compared to CUDA cores~\cite{nvidia_a100}, we achieve a lower speedup on average because we rely on operations using CUDA cores. As described in Section~\ref{sec:euclidean_tensor}, combining CUDA and TCs to compute FP64 Euclidean distances is required due to the restricted matrix sizes when using the WMMA API and FP64.

Compared to the multi-core CPU algorithms \superego~\cite{superego} and \hilbert~\cite{fgfHilbert}, we find that \tensor typically outperforms these algorithms.


 Future work includes, investigating cache and shared memory efficiency, particularly for higher dimensions,  modeling TC performance to determine in which scenarios they should be leveraged instead of CUDA cores, using other floating point precisions available for TCs, and incorporating our TC GPU kernel into other algorithms, such as  $k$NN~\cite{gowanlock_knn}, and particle simulations such as those in molecular dynamics~\cite{molecularDyn}.

\section*{Acknowledgements}
This material is based upon work supported by the National Science Foundation under Grant No. 2042155.

\bibliographystyle{IEEEtran}
\bibliography{main.bib}

\end{document}